\documentclass[11pt,a4paper,english]{article} 

\usepackage{soul}
\usepackage{natbib}
\usepackage{url}
\usepackage{comment}
\usepackage{amsmath}
\usepackage{lineno}
\usepackage{cancel}
\usepackage{mdframed}
\usepackage{ragged2e}
\usepackage{xcolor}
\usepackage{multicol}
\usepackage[stable]{footmisc}
\usepackage[centering,noheadfoot,margin=0.75in]{geometry}

\definecolor{pink}{HTML}{EE2967}

\renewenvironment{quote}
  {\list{}{\rightmargin=17pt \leftmargin=17pt}%
   \item[]}%
  {\endlist}

\AtBeginDocument{%
  }

\begin{document}

\title{In Defense of the Turing Test and its Legacy\thanks{This article develops the author's contribution to the session ``The Turing Test and its Legacy, 1950-2025'' at The Next Turing Tests Conference, held in October 2025 at the University of Cambridge.}}

\author{Bernardo Gonçalves\vspace{10pt}\\National Laboratory for Scientific Computing (LNCC), Brazil\vspace{1pt}\\goncalves@lncc.br}
\date{November 20, 2025}
\maketitle

\vspace{-18pt}
\begin{abstract}\noindent
Considering that Turing's original test was co-opted by Weizenbaum and that six of the most common criticisms of the Turing test are unfair to both Turing's argument and the historical development of AI. 

\end{abstract}
\vspace{-5pt}

\vspace{15pt}

\noindent
The Turing test has faced criticism for decades, most recently at the Royal Society event ``Celebrating the 75th Anniversary of the Turing Test.'' The question of the Turing test's significance has intensified with recent advances in large language model technology, which now enable machines to pass it. In this article, I address six of the most common criticisms of the Turing test: 

\begin{itemize}
\item The Turing test encourages fooling people; 
\item Turing overestimated human intelligence, as people can be easily fooled (the ELIZA effect);
\item The Turing test is not a good benchmark for AI; 
\item Turing's 1950 paper is not serious and/or has contradictions;
\item Imitation should not be a goal for AI, and it is also harmful to society; 
\item Passing the Turing test teaches nothing about AI. 
\end{itemize}

\noindent
All six criticisms largely derive from Joseph Weizenbaum's influential reinterpretation of the Turing test. The first four fail to withstand a close examination of the internal logic of Turing's 1950 paper, particularly when the paper is situated within its mid-twentieth-century context. The fifth also arises from a naive view that overlooks the political-economy context and how AI is imbricated in the broader history of automation in modern societies. The sixth also arises partly from ``the AI effect'' (discussed below) and partly from the dynamics of polarized debates.

\section*{Does Turing's original test encourage fooling people?}

Turing started using the term \emph{imitation} after his wartime experience breaking the Enigma codes by studying the machine's behavior alone. If its actions could be predicted with computing machines, he wondered, might computers not be capable of ``imitating'' the brain? 

Turing’s proposed question was whether a machine could \emph{learn} to ``play the imitation game so well'' that it would be mistaken by ordinary people for what it is not. He framed the problem this way because some of his closest critics insisted that machines would never be capable of mastering language, a capacity they viewed as the hallmark of thinking \citep{goncalves2024beautiful}. A decade later, Weizenbaum shifted the focus of this debate from the possibility of machine intelligence to the (no less important) issue of human susceptibility to being misled by computing technologies.

Weizenbaum's question appeared as the title of a paper he published in 1962 in the \emph{Datamation} magazine, ``How to Make a Computer \emph{Appear} Intelligent'' \cite[his emphasis]{weizenbaum1962}. 
At first sight, Turing's and Weizenbaum's questions may seem remarkably similar. To see the difference more clearly, note that Weizenbaum had already achieved his main result by 1966 \citep{weizenbaum1966} --- just four years later. Turing, by contrast, envisioned a fifty-year project that ultimately took more than seventy.
These differing time frames are revealing. Turing sought to explore whether machines could learn enough to imitate what was widely accepted as human intelligence. Weizenbaum, by contrast, bypassed the learning question altogether: he set out to construct the simplest possible computer, pre-scripted with psychological prompts, to test whether people would attribute intelligence to it. 

Did Turing exclude Weizenbaum-style machines from his test? Yes. This is evident, for example, in his discussion of the ``human fallibility'' he encouraged the machine to exhibit. Fallibility was intended to emerge as a by-product of learning from experience, rather than from trickery:

\begin{quote}\noindent
Another important result of preparing our machine for its part in the imitation game by a process of teaching and learning is that `human fallibility' is likely to be omitted [from the teaching] in a rather natural way, i.e., [learned] without special `coaching.''' 
\cite[p.~459]{turing1950}
\end{quote}

\noindent
A machine qualifies for Turing's test only if it was \emph{not} engineered specifically to pass it --- a condition that excludes Weizenbaum's interpretation of the test. Otherwise --- and this is a crucial point --- the situation would be akin to a physicist having to check for sabotage before running an experiment.

Further clarification came in Turing’s 1951 lecture, ``Can Digital Computers Think?'':

\begin{quote}\noindent
This [learning] process could probably be hastened by a suitable selection of the experiences to which it was subjected. This might be called `education.' But here we have to be careful. It would be quite easy to arrange the experiences in such a way that they automatically caused the structure of the machine to build up into a previously intended form, and this would obviously be a gross form of cheating, almost on a par with having a man inside the machine. 
\citep{turing1951b}
\end{quote}

\noindent
This shows that Turing did not encourage cheap deception or psychological manipulation. Yet these were precisely the strategies that Weizenbaum later adopted as a methodological stance. In Turing's test, in contrast, deception functioned merely as a playful means to an end.

\vspace{-4pt}
\section*{The ELIZA effect: Did Turing overestimate human intelligence?}
\vspace{-4pt}

Contrary to recent opinions and headlines, there is direct textual evidence that Turing was aware of the phenomenon that would become known as ``the ELIZA effect'' --- namely, that many people can easily be fooled into imagining intelligence in a computer's behavior, or anthropomorphizing it. 
In his 1948 report ``Intelligent Machinery,'' Turing notes: 

\begin{quote}\noindent
Playing [chess] against such a machine gives a definite feeling that one is pitting one's wits against something alive. 
\citep{turing1948}
\end{quote}

\noindent
The report ends with a section called ``Intelligence as an Emotional Concept,'' where he writes: 

\begin{quote}\noindent
The extent to which we regard something as behaving in an intelligent manner is determined as much by our own state of mind and training as by the properties of the object under consideration. If we are able to explain and predict its behaviour or if there seems to be little underlying plan, we have little temptation to imagine intelligence. With the same object therefore it is possible that one man would consider it as intelligent and another would not; the second man would have found out the rules of its behaviour.
\citep{turing1948}
\end{quote}

\noindent
This shows that, by 1948, Turing was already prepared for the insights that would later emerge from Weizenbaum's work in the 1960s. One might ask, then, why Turing made the interrogators in his test ``average'' people \cite[p.~442]{turing1950} --- or as he later reiterated, ``not be expert about machines.''

One reason is that Turing regarded experts as biased in the opposite direction. As he observed in the passage above, ``we have little temptation to imagine intelligence'' in a machine if we can ``explain and predict its behaviour or if there seems to be little underlying plan.'' He was thus prepared to expect experts ``explaining away'' the possibility of intelligence in machines --- a phenomenon closely related to what would later be called ``the AI effect,'' the conceptual dual of the ELIZA effect. The AI effect is the tendency to assume a task requires intelligence when a machine fails to perform it, yet subsequently reclassify the task as \emph{not} requiring intelligence once the machine succeeds.

Another reason is that by making ordinary people the interrogators and judges, Turing ensured that the test-passing moment would mark a broader social and cultural shift: the moment when public perception of machines had changed. In Turing's words, ``the use of words and general educated opinion will have altered so much that one will be able to speak of machines thinking without expecting to be contradicted'' \cite[p.~442]{turing1950}. 
Experts might be too close to notice it.

\section*{Is the Turing test not a good benchmark for AI?}
\noindent
AI pioneers such as John McCarthy and Marvin Minsky never contemplated conducting practical Turing tests. Instead, they received Turing's proposal as a ``definition,'' a ``strong criterion,'' and a conceptual foundation for the AI problem: ``that of making a machine behave in ways that would be called intelligent if a human were so behaving'' \cite[p.~46]{goncalves2024beautiful}. 
In 1967, in \emph{Computation: Finite and Infinite Machines}, Minsky emphasized section \S6 of Turing's paper: ``Turing discusses some of these issues in his brilliant [1950] article $\hdots$ and I will not recapitulate his arguments $\hdots$ They amount, in my view, to a satisfactory refutation of many such objections.'' 

However, in 1966, Weizenbaum described his ELIZA experiment as a ``striking form of Turing's test'' \cite[p.~42]{weizenbaum1966}. The impact of his reinterpretation of the test would become evident from the early 1990s onward with the emergence of practical ``Turing'' tests. As known, entrepreneur Hugh Loebner established a prize for chatbots in Turing’s name. Weizenbaum's influence was unmistakable: in the first three editions, the winning program was the ``PC Therapist,'' created by Joseph Weintraub, a psychology graduate turned computer programmer. Although it was common sense that no Turing-test-passing computer program existed at the time, Loebner’s initiative nonetheless capitalized on Turing's legacy. The annual competition continued through the 1990s into the 2010s, consistently attracting Weizenbaum-style chatbots and even sponsorship from the Royal Society of London. 

In 1995, Minsky urged Hugh Loebner to ``revoke his stupid prize, save himself some money, and spare us the horror of this obnoxious and unproductive annual publicity campaign.'' That same year, Patrick Hayes and Kenneth Ford delivered their talk ``The Turing Test Considered Harmful'' at IJCAI \citep{hayes1995}. 
Note that this is a high moment of the Loebner campaign. By then, the notion of practical ``Turing'' tests had already come to dominate the prevailing understanding of the Turing test concept. Hayes and Ford's paper provides telling evidence of this shift. While the authors claimed to take Turing's 1950 paper seriously, and indeed offered a compelling critical analysis of the imitation tests viewed as practical experiments, their reading was fundamentally anachronistic. They sought the design of a concrete, practical experiment rather than engaging with the text on its own terms. The critical question, therefore, is: What, then, is actually in Turing's text?


Turing’s paper presents a thought experiment illustrating the true power of digital computers, showing that they could be more than mere advanced calculators \citep{goncalves2024beautiful}. Spanning 28 pages and seven sections (\S1--\S7), a close structural reading reveals three main parts or logical steps: 

\begin{itemize}
\item \emph{The Proposal} (sections \S1-\S3, in 3+ pages), where Turing introduces the imitation game, discuss its validity and limitations, and the kinds of machines eligible to take part in it; 
\item \emph{The Science} (sections \S4 and \S5, in 6 pages), where Turing presents the emerging science and technology of digital computing and explains the universality property of digital computers grounded in the universal Turing machine concept;
\item \emph{The Discussion} (sections \S6 and \S7, in 18+ pages), where Turing formulates and answers nine objections to the possibility of machine intelligence and outlines a research program to build learning machines that could pass his test. 
\end{itemize}

\noindent
This structure --- with its thematic organization and unbalanced distribution --- offers insight into the nature of Turing's argument. Across all three main logical steps of the text, Turing \emph{discursively} presents several variants of his test. This structural feature of the paper has, at least since Hayes and Ford's critique, often been regarded as a flaw. 
Yet the presence of multiple variants of the test instead suggests that the imitation game functions as an illustrative conceptual device within Turing's broader argument, rather than as an end in itself. This is evident, for instance, in section \S6 (the longest of the paper), where he explicitly returns to the imitation game while responding to several objections. 
The fundamental question Turing poses is whether A could imitate the intellectual stereotypes associated with B well enough to deceive C, despite their natural differences.

Under this perspective, the constant variation in its instantiated conditions is not a flaw, but a feature vividly illustrating his concept of universal imitation. 
By applying universality (part 2, \emph{the science}) to the imitation game (part 1, \emph{the proposal}), Turing indicated, early on, the power of digital computing (part 3, \emph{the discussion}), showing how digital processes could blur the boundaries between natural and social types to the point of practical indistinguishability. 
We have reached this juncture, with the Turing test having served as an early warning and as a crucial conceptual and historical benchmark that guides our understanding of AI's development.

\section*{Is Turing's 1950 text not serious and does it have contradictions?}

Turing’s 1950 paper is remarkable not only for its accessibility and engaging style but also for its status as one of the most cited and foundational texts in modern philosophy. Turing had a sense of humor, but he used irony purposefully --- as a \emph{method} \cite[Ch. 7]{goncalves2024argument}. He continued an English tradition of irony that has been described, for example, in relation to the philosophical style of David Hume: ``Irony gave him a method of operating in a world that found his ideas both strange and shocking." 

The various forms of the test, rather than reflecting imprecision and poor design choices as argued by Hayes and Ford (\citeyear{hayes1995}), can be seen instead as neat concessions to Turing's critics. For instance, Hayes and Ford considered Turing's focus on ``human conversational competence'' to be species-biased and circular, as there is no definition of that quality other than to ``pass the corresponding Turing test.'' However, the demand for mastery of human language actually came from Turing's critics, namely Geoffrey Jefferson and Michael Polanyi \cite[Ch.~4]{goncalves2024argument}. Before their objections, raised between June and December 1949  \citep{goncalves2024beautiful}, Turing was happy with machine chess as an initial testbed for machine intelligence. His interlocutors, however, dismissed chess as unimpressive on the grounds that it was rule-bound. 
Hayes and Ford also argued that ``The gender test is not a test of making an artificial human, but of making a mechanical transvestite.'' However, by including role B (the assistant) in a gender test, Turing was mirroring a thought experiment proposed by Jefferson just months earlier. Jefferson had considered placing one of Grey Walter's mechanical turtles alongside a natural one. He expected that the natural turtle would expose the artificiality of the mechanical one, arguing that gendered behavior is causally linked to the physiology of male and female sex hormones \citep{goncalves2024beautiful}.
Turing was compelled to respond. 
Overall, the Turing test's various designs conform to Karl Popper's rule for using imaginary experiments in critical argumentation: ``the idealizations made must be concessions to the opponent, or at least acceptable to the opponent'' \cite[Ch.~5]{goncalves2024argument}. 

Turing states that the main question --- can machines think? --- is ``too meaningless to deserve discussion.'' Yet, the longest section of his paper, section \S6, is titled ``Contrary Views on the Main Question,'' and he discusses it in detail. This is often seen as a contradiction. However, a close reading shows that Turing is not actually engaging with the main question directly. Rather, he presents nine objections and responds to them \emph{through} the framework of the imitation game, which he uses to reframe the main question, placing machines and humans on an equal footing.

We are now in a position to examine the method Turing used to structure his philosophical argument in the third logical step of his paper (sections \S6 and \S7, \emph{the discussion}). This method is the Socratic dialectic \cite[Ch.~6]{goncalves2024argument}, as described by Bertrand Russell at the end of his chapter on Socrates in \emph{A History of Western Philosophy}, one of the few works Turing cites. 
Russell argues that it ``tends to promote logical consistency, and is in this way useful $\hdots$ But it is quite unavailing when the object is to discover new facts.'' Turing followed this principle strictly, writing: 

\begin{quote}\noindent
The reader will have anticipated that I have no very convincing arguments of a positive nature to support my views $\hdots$ The only really satisfactory support that can be given for the view expressed at the beginning of \S6, will be that provided by waiting for the end of the century and then doing the experiment described. \cite[pp.~454-455]{turing1950}
\end{quote}

\noindent
Turing's rhetoric of ``doing the experiment'' can be best understood as the making of a learning machine (A) capable of answering unrestricted questions indistinguishably from a chosen human type (B), as if participating in one of his imitation games. 
Turing used dialectic to discuss the original question as a ``logical'' question in Russell’s sense (\S6), and proposed a long-term research project to address it as an ``empirical'' question (\S7), both framed through the imitation game.

\section*{Has the emphasis on imitation been harmful to AI and society?}

\noindent
Machine-based imitation can be traced back at least to nineteenth-century knitting frames \citep{merchant2023}, which replicated the work of textile craftsmen more cheaply and quickly --- though at the cost of fossil-fuel energy consumption --- to produce fabrics that appeared indistinguishable from handcrafted ones to the ordinary observer (cp. the logic of the Turing test). Although it still required a human operator, the machine could produce output far beyond anything individual craftsmen could achieve. One might say that the operator was ``augmented'' by the frame, but such a view overlooks the several workers whose livelihoods were displaced by the machine. 

Now it is the time of intelligence, as Turing warned. For this, he has recently been associated with ``the Turing trap'' \citep{brynjolfsson2022}: ``As machines become better substitutes for human labor, workers lose economic and political bargaining power and become increasingly dependent on those who control the technology.'' Turing is criticized for framing the goal of AI as imitating --- and potentially substituting --- humans rather than augmenting them. The paradigmatic illustration given to that distinction is Jeff Bezos' alleged augmentation --- rather than automation --- of bookstores (p.~280). Do Amazon's subcontracted workers have any bargaining power? ``The Turing trap'' obscures the possibility of democratic control over the technology. Further, consider an Amazon warehouse: it is unclear whether machines are augmenting workers’ capabilities or the reverse. In any case, how can we know that the residual human skills augmented by machines will not soon become obsolete?  
The oft-invoked automation-augmentation distinction remains conceptually vague, empirically undertested, and only weakly justified in light of economic history. 

The Turing-inspired definition of the AI problem by McCarthy, Minsky and others provided the flexibility needed for the field to evolve coherently across techniques and tasks, even as the notion of what constitutes an ``intelligent'' task continually shifted with the AI effect. This owes much to the political-economic incentives for automation that have long been deeply embedded in modern societies. Should the mid-twentieth-century Turing test be held responsible for the societal issues we face today? Why not, instead, question the socioeconomic model that concentrates the wealth generated by automation at the expense of the common good? Why is the automation of white-collar work contested, while the automation of blue-collar work is often treated as acceptable?

\section*{The scientific significance of passing the Turing test}

The current approach to AI poses risks to both nature and society. Yet passing the Turing test has yielded at least two significant scientific discoveries. First, machine learning systems exhibit the critical-scale phenomena conjectured by Turing (\citeyear[p.~454]{turing1950}): once a model surpasses certain thresholds of data and compute, characteristic improvements in generalization emerge. These dynamics deserve further study, despite the tendency toward over-interpretation. Second, human language can be effectively mastered through distributional semantics, i.e., by deriving meaning from patterns of use rather than from hand-crafted rules. Before this achievement, who would have claimed that language learning and use require no intelligence?

\noindent
\newpage

\section*{Acknowledgments}
The author is indebted to Moshe Vardi for his valuable comments on an earlier draft of this paper. The author alone is responsible for the accuracy of this article.
This research has been supported by the Carlos Chagas Filho Research Support Foundation for the State of Rio de Janeiro (FAPERJ grant E-26/210.511/2025). 

\bibliographystyle{ACM-Reference-Format}
\bibliography{turing-futures-sep2024}

\end{document}